\documentclass[prd,aps]{revtex4}
\usepackage{graphicx}
\newcommand{\bb}{\begin{eqnarray}}
\newcommand{\ee}{\end{eqnarray}}

\begin{document}
\title{ \bf Electron  bound by a potential well in the presence of
a constant uniform magnetic field}
\author{V. R. Khalilov$^1$ and F. Kh. Chibirova$^2$}
\affiliation{$^1$Faculty of Physics, Moscow State University,
119899, Moscow, Russia\\ $^2$Karpov Institute of Physical
Chemistry, 103064, Moscow, Russia}

\begin{abstract}
We study the effect of a constant uniform magnetic field on an
electrically charged massive particle (an electron) bound by a
potential well, which is described by means of a single attractive
$\lambda\delta({\bf r})$ potential. A transcendental equation that
determines the electron energy spectrum is derived and solved. The
electron wave function in the ground (bound) state is
approximately constructed in a remarkable simple form. It is shown
that there arises the probability current in the bound state in
the presence of a uniform constant magnetic field. This (electric)
current, being by the gauge invariant quantity, must be observable
and involve (and exercise influence on) the electron scattering.
The probability current density resembles a stack of ``pancake"
vortices'' whose circulating ``currents'' around the magnetic
field direction ($z$-axes) are mostly confined within the plane
$z=0$. We also compute the tunnelling probability of electron from
the bound to free state under a weak constant homogeneous electric
field, which is parallel to the magnetic field. The model under
consideration is briefly discussed in two spatial dimensions.

\end{abstract}

\pacs{PACS numbers: 03.65.Nk, 03.65.Ge, 03.65.-w}

\maketitle

\section{Introduction}
The behavior of quantum nonrelativistic systems in external
electromagnetic fields has attracted permanent interest in view of
possible applications of the corresponding models in many
phenomena of quantum mechanics. Bound electron states play an
important role in quantum systems in condensed matter. When the
external field configuration has the cylindrical symmetry a
natural assumption is that the relevant quantum mechanical system
is invariant along the symmetry (for example, $z$) axis and the
quantum mechanical problem then becomes essentially
two-dimensional in the $xy$ plane. Such is the case when the
external field configuration is a superposition of a constant
uniform magnetic field and a cylindrically symmetric potential.
Nonrelativistic electrons in such external field backgrounds are
good quantum mechanical models for studying remarkable macroscopic
quantum phenomena such as the fractional quantum Hall effect$^1$
and high-temperature superconductivity$^2$. A related problem is
the behavior of electrons in trapping potentials in the presence
of a constant uniform magnetic field in an effective mass
approximation. Magnetic fields seem to be likely to effect on
weakly bound electrons into singular potentials of defects in the
defect films$^3$ and solids$^4$.

Pure two-dimensional models also are of significant interest. The
effect of magnetic fields on a weakly bound electron in two
spatial dimensions was studied by us in Ref. 5. This model is of
interest because it gives a good example of a nonrelativistic
analog of the so-called dimensional transmutation phenomenon first
discovered by S. Coleman and E. Weinberg in Ref.\cite{colw} in the
massless scalar electrodynamics. Whereas above nonrelativistic
systems can be described by the Schr\"odinger equation,
relativistic systems, related to the Dirac Hamiltonian in 2+1
dimensions in a constant uniform magnetic field, show up in a
certain type of doped two-dimensional semimetals$^7$. Similar
problems are also related to a number of problems in quantum
theory, for example, the parity violation, the theory of anyons
(particles satisfying a fractional statistics), the Aharonov--Bohm
effect$^8$, and other. At last recently, a new type of spectral
problem has been found$^{9,10}$ in quantum mechanics of planar
electrons in a superposition of constant uniform magnetic and
cylindrically symmetric potential fields. For this new class of
spectral problems, the so-called quasi-exactly solvable (QES)
models the energy spectra exist if a certain relation between the
parameters characterizing the intensity of the interaction of an
electron with external fields hold and then  solutions of the
corresponding equations of quantum mechanics can be expressed as
the product of a weight function and a finite polynomial. The some
physical examples of QES models, which include the two-dimensional
Schr\"odinger or Dirac equation for an electron in a superposition
of a constant uniform magnetic field and an attractive Coulomb
field, were studied in Ref. \cite{hokh}.

  The purpose of this paper is to study the effect of a constant
uniform  magnetic field on an electron bound by  a single
attractive $\lambda\delta({\bf r})$ potential. We derive a simple
transcendental equation determining the electron energy spectrum
and construct the approximate wave function for a bound electron
state in the presence of a constant uniform magnetic field. We
show that the sizes of the electron localization region  change
and the probability current arises even when the electron is in
the bound state in a superposition of a constant uniform magnetic
field and a single attractive $\lambda\delta({\bf r})$ potential.
The probability current in three-dimensional space resembles a
stack of ``pancake vortices'' whose circulating (around the
$z$-axes) ``currents'' are mostly confined in the weak magnetic
field within the plane $z=0$.

The equation for determining the energy levels of the electron
states is also obtained for the model under study in two spatial
dimensions. We show that in difference from the three-dimensional
case the binding energy is not analytical in $\lambda$.

The tunnelling of electron from the bound to free state under a
weak electric field is of importance. In present paper we compute
the tunnelling probability of electron from the bound to free
state under a weak constant homogeneous electric field, which is
parallel to the magnetic field.

\section{Electron in an Potential Well in the Presence of a Constant Uniform
Magnetic Field}

Let us consider an electron with the charge $e<0$ in an attractive
singular potential of the form \bb U(r)=
-\frac{\hbar^2\lambda}{2m}\delta({\bf r}) \label{well1} \ee and a
constant uniform magnetic field ${\bf B}$, which is specified in
Cartesian coordinates as \bb
 {\bf B}=(0,\,0,\,B)=\nabla\times {\bf A},\ \
 {\bf A}=(-yB,\,0,\,0)
\label{e1} \ee In (\ref{well1}) $\lambda$ is a positive coupling
constant of the length dimension, $\delta({\bf r})$ is the
three-dimensional Dirac delta function, $m>0$ is the effective
mass of an electron. It is well to note that the attractive
$\lambda\delta({\bf r})$  potential can be considered as the limit
of sequence of appropriate narrow rectangular potential wells \bb
U(r)= =-U_0,\quad r<R,\qquad U(r) = 0,\quad r>R, \label{well} \ee
then the parameter $\lambda$ is expressed via $U_0$ and $R$ as
$$
\lambda=\frac{2mU_0R^3}{\hbar^2}.
$$

The Schr\"odinger equation is \bb
\frac{1}{2m}\left[\left(-i\hbar\frac{\partial}{\partial
x}+\frac{eB}{c}y\right)^2 -\hbar^2\frac{\partial^2}{\partial y^2}
-\hbar^2\frac{\partial^2}{\partial z^2}-\hbar^2\lambda\delta({\bf
r})\right] \Psi_E({\bf r})=E\Psi_E({\bf r}). \label{e21} \ee

The electron wave function in magnetic field (\ref{e1}) can be
found in the form$^{12}$ \bb
 \psi_{np}(t, {\bf r})=\frac12 e^{-iE_{ns}t/\hbar}e^{i(p_1x+p_3z)/\hbar}V_n(Y)
\left( \begin{array}{c}
1+s\\
1-s
\end{array}\right)
\label{sol1} \ee where \bb E_{ns}=\hbar \omega
\left(n+\frac12\right)+\frac{p_3^2}{2m} +s\hbar
\omega\frac{m}{2m_e} \label{e2} \ee is the energy eigenvalue,
$\omega=|eB|/mc$, $m_e$ is the mass of free electron, $p_1, p_3$
are the eigenvalues of generalized momentum operator and  $s=\pm
1$ is the spin quantum number. Note that for $m_e=m$ and for ${\rm
sign} eB<0$, all the energy levels except one with $n=0,\quad
s=-1$ are doubly degenerate: the coincident levels are those with
$n \quad s=1$ and $n+1,\quad s=-1$. In given paper we consider the
case $m_e=m$. Then, the energy eigenvalues depend only on the
number $n$.

Note that $p_1$ is
constrained by $|p_1|\leq eBL/c$ (see, Ref. \cite{ll}). The
functions
$$
 V_n(Y) = \frac{1}{(2^n n!\pi^{1/2}a)^{1/2}}
\exp\left(-\frac{(y-y_0)^2}{2a^2}\right)H_n\left(\frac{y-y_0}{a}\right),
$$
are expressed through the Hermite polynomials $H_n(z)$, the
integer $n=0, 1, 2, \dots$ indicates the Landau level number,
$$
a=\sqrt{\frac{\hbar c}{|eB|}}\equiv \sqrt{\frac{\hbar}{m\omega}}
$$ is the so-called magnetic length and $y_0=-cp_1/eB$. It should be
reminded that the classical trajectory of electron  in the $xy$
plane is a circle. The quantity $y_0$ corresponds to the classical
$y$ - coordinate of the circle center. All the electron states
(\ref{sol1}) are not localized in the $x, z$-directions.

Solutions of Eq.(\ref{e21}) are sought in the form
\bb \Psi_E({\bf r})=\sum\limits_{n=0}^{\infty} \int dp_1dp_3
C_{Enp}\psi_{np}({\bf r}) \equiv \sum\limits_{n,p}
C_{Enp}\psi_{np}({\bf r}), \label{sum1} \ee where $\psi_{np}({\bf
r})$ is the spatial part of wave functions (\ref{sol1}).

Then, for coefficients $C_{Enp}$,  one obtains \bb
C_{Enp}(n+cp_3^2+b)=\lambda_0\sum\limits_{l,k}C_{Elk}V_l(0)V_n(0),
\label{syst}\ee where
$$
c=\frac{1}{2m\hbar\omega}, \quad b=-\frac{E}{\hbar\omega}, \quad
\lambda_0=\frac{\lambda}{8\pi^2m\hbar\omega},\quad V_l(0)\equiv
V_l(y=0).
$$
Note that terms on the right of Eq.(\ref{syst}) are the matrix
elements of the interaction operator of electron with the singular
potential (\ref{well1}). Correctly these matrix elements must be
obtained at first for the interaction potential (\ref{well}) with
finite $U_0$,  $R$ and then by proceeding to the limits $R\to 0$,
$U_0\to\infty$. But for finite $U_0$ and $R$  the main
contribution in the matrix element
$U_{np_3,n^{\prime}p_3^{\prime}}$ gives the quantum number regions
$|p_3-p_3^{\prime}|R<\hbar$, $\sqrt{n-n^{\prime}}R<a$ because the
matrix element falls off quickly due to the oscillations outside
the above regions. Moreover, for the sharp potentials, one can
obtain $U_{np_3,n^{\prime}p_3^{\prime}}\sim -\lambda$ for  $n,
n^{\prime}<N$ and $p_3, p_3^{\prime}<P$ and
$U_{np_3,n^{\prime}p_3^{\prime}}=0$ for $n, n^{\prime}>N$ and
$p_3, p_3^{\prime}>P$,  $\lambda\to 0$ $N, P\to\infty$ as $R\to
0$. So, for such potentials the sum taken over $n$ on the right of
Eq. (\ref{syst}) is limited by the value $N$ and the bound state
energy becomes depending on $N$ and $\lambda$, which, as well as
$U_0$ and $R$, are considered finite. In the limits $R\to 0$,
$U_0\to\infty$, according to above given estimations, we have
$\lambda\to 0$ $N\to\infty$.

Thus, when transforming to the singular potential  the bound state
energy remains finite (and, moreover, any given) value only if $N$
will tend to infinity as \bb
N^{3/2}=\frac{12\sqrt{2}a\pi}{\lambda}+
\left(-\frac{E_0}{\hbar\omega}\right)^{3/2} \label{toN} \ee as
$R\to 0$.

Let us write  $C_{Enp}$ as
 \bb C_{Enp}=C_E\frac{V_n(0)}{n+cp_3^2+b}
\label{sup}\ee, insert (\ref{sup}) in (\ref{syst})and take account
of the formulas \bb \sum\limits_{n,p}
|C_{Enp}|^2=1,\label{relat}\ee \bb \int
dp_1V_n(0)V_k(0)=\frac{\hbar}{a^2}\delta_{n,k}.\label{import}\ee
Then one obtains equations \bb
1=\frac{\lambda}{8\pi^2m\hbar\omega}\sum\limits_{n=0}^{N}
\int\limits_{-\infty}^{\infty}dp_3\frac{1}{n+b+cp_3^2},
\label{ener}\ee \bb (C_E)^{-2}=
\sum\limits_{n=0}^{N}\int\limits_{-\infty}^{\infty}dp_3
\frac{1}{(n+b+cp_3^2)^2}. \label{coeff}\ee Integrating the right
side of Eq.(\ref{ener}) we obtain the following transcendental
equation \bb
1=\frac{\lambda\sqrt{m}}{8\sqrt{2}\pi}\sum\limits_{n=0}^{N}
\frac{\omega}{\sqrt{|\hbar\omega n -E|}}, \label{energy1}\ee the
roots of which determines the energy levels of electron in the
considered combination of fields. This equation can be solved
numerically (graphically); its roots $x_n=E_n/\hbar\omega$ for
different $\lambda$ is determined by the crossing of horizontal
$8\sqrt{2}a\pi/\lambda$ with different values $\lambda$ with the
function \bb f(x)=\sum\limits_{n=0}^{N} \frac{1}{\sqrt{|n-x|}}.
\label{rootener}\ee

It is seen from Eq. (\ref{rootener}) that for attractive potential
the main contributions in the sum gives small lengths $\Delta x$,
which adjoins to the corresponding root with $n$ to the left. Plot
of $f(x)$ as a function of $x$ near the roots with $n=1, 2, 3, 6,
10$ is given in  Fig. 1 for $N=10^6$.

\begin{figure}[h]
\vspace{-0.2cm} \centering
\includegraphics[angle=270, scale=0.5]{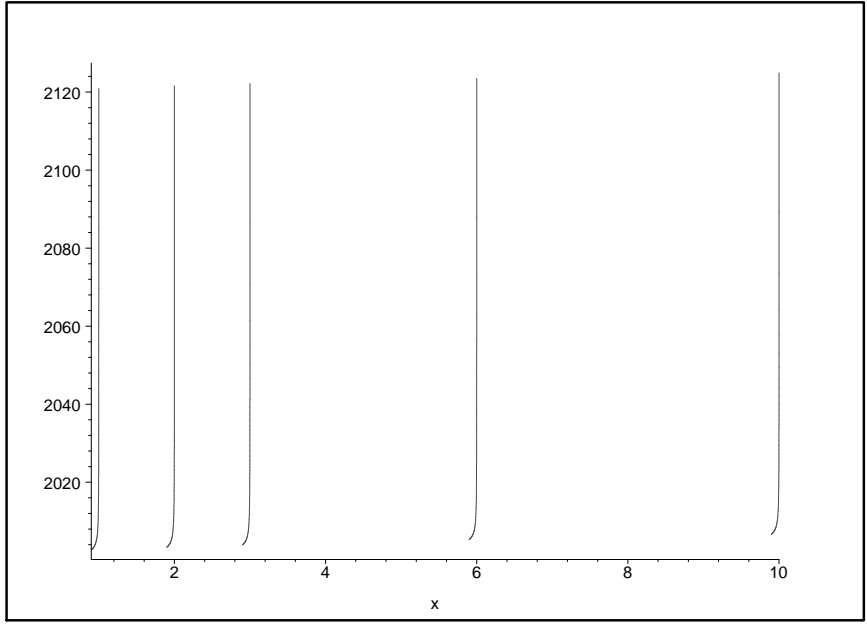}
\caption{Plot of the function $f(x)$} \vspace{-0.1cm}
\end{figure}

It follows from Eq. (\ref{energy1}) that if the potential energy
is small compared with $\hbar\omega$ then the energy level $E_n$
is located between two energy levels $\hbar\omega n$  and
$\hbar\omega(n+1)$ with the exception of the negative energy level
$E_0$.

The energy levels can be approximately calculated if
$$
\frac{\lambda}{8\pi a\sqrt{2}}\ll 1.
$$
Then, representing \bb E_n=\hbar\omega n+\delta_n,\quad \delta<0
\label{energy2}\ee for any number  $n=0, 1, \ldots$, we obtain \bb
\delta_n=-\frac{\lambda^2m\omega^2}{32\pi^2}. \label{energy3}\ee
First, let the singular potential can be considered as a
perturbation therefore the coupling constant $\lambda<a$. All the
levels $\delta_n$  are shifted down with respect to the
corresponding Landau levels   $\hbar\omega n$ for any $n=0, 1,
\ldots$. In this case $|\delta_n|\ll \hbar\omega$. If $E_0<0$ is
the electron energy in the pure singular $\lambda\delta({\bf r})$
potential then $\delta_0$ is the magnetic field correction to the
level $E_0$.

If the coupling constant $\lambda$ obeys the inequality
$8\pi\sqrt{2}\gg\lambda>a$ then  $|E|>\hbar\omega$ and $E$ is
determined by Eq. (\ref{energy3}). It should be emphasized there
exists the only energy level ($E_0<0$) in a single attractive
$\lambda\delta({\bf r})$ potential.

Plot of $f(x)$ as a function of $x$ near the root with $n=0$,
which corresponds ground (bound) state is given in Fig. 2 for
$N=10^6$.

\begin{figure}[h]
\vspace{-0.2cm} \centering
\includegraphics[angle=270, scale=0.5]{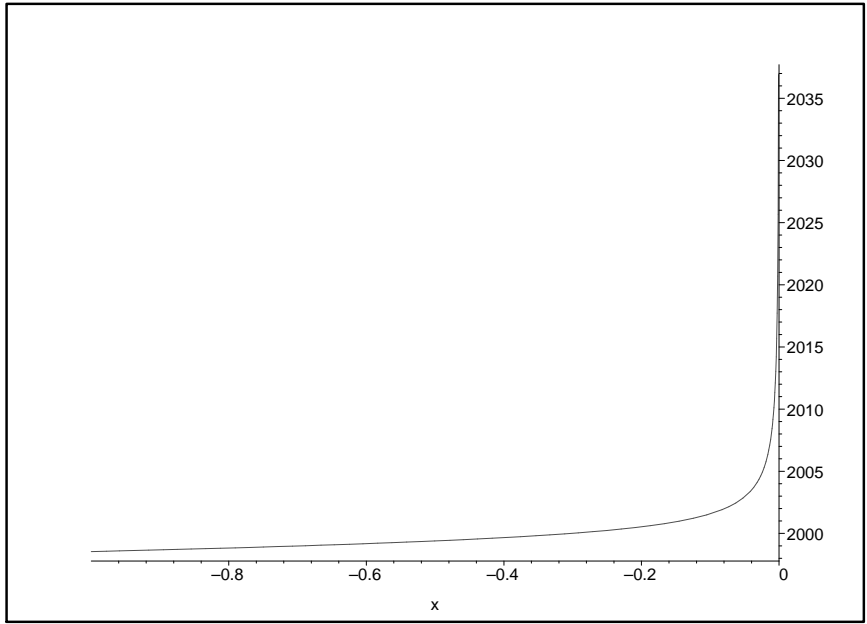}
\caption{Function $f(x)$ near the root $n=0$} \vspace{-0.1cm}
\end{figure}

Graphically the root $x_0=E_0/\hbar\omega$ is determined by the
crossing of horizontal $8\sqrt{2}a\pi/\lambda$ with the function
plotted in Fig. 2. We also see that the energy of bound electron
state is analytical in $\lambda$.

The electron energy in bound state is not analytical in $\lambda$
for the model under discussion in the two-dimensional case.
Indeed, Eq.(\ref{energy1}) should be replaced by the equation \bb
1=\frac{\lambda}{4\pi}\sum\limits_{n=0}^{N}\frac{1}{n+b},\quad
b=-\frac{E}{\hbar\omega} \label{ener3}\ee for the two-dimensional
case (see, \cite{chib}). Here we took account of all remarks
concerning the number $N$, which were given when deriving of Eq.
(\ref{energy1}). The energy of ground state (that is a negative
root of Eq.(\ref{ener3}) $E_0<0$) can be easily found for $N\gg
1$. Replacing the summation over $n$ by the integration, we find
as a result \bb
1=\frac{\lambda}{4\pi}\ln\left(\frac{N}{b}\right)\label{cut} \ee
and for the electron energy of bound state we obtain
 \bb
E_0=-\hbar\omega N\exp\left(-\frac{4\pi}{\lambda}\right).
\label{binde} \ee

It is of importance to emphasize that in the two-dimensional case
the coupling constant $\lambda$ is the dimensionless constant.
Nevertheless, there exists the bound state in the attractive
$\lambda\delta({\bf r})$ potential. In the limits $N\to \infty$,
$\lambda\to 0$ we must require that $N$ should depend on the
dimensionless constant $\lambda$ so as the binding energy $-E_0$
would remain finite as $N\to \infty$. Thus, the cutoff
dimensionless parameter $N$, which tends to infinity, transmutes
in arbitrary binding energy $|E_0|$. This is the nonrelativistic
analog the dimensional transmutation phenomenon. For the model
under discussion in the absence of magnetic field this phenomenon
was first considered in Ref.\cite{thor}.

\section{Electron Wave Function and Probability Current in the Bound State}

Now we shall construct the wave function of electron in the
negative-energy state with $n=0$. It can be found from
Eq.(\ref{sum1}) by putting $n=0$. Then,  using Eqs.(\ref{sup}),
(\ref{coeff}), we integrate over $p_1$ by means of the following
integral$^{14}$ \bb \int\limits_{-\infty}^{\infty}dx
e^{-ixy}U_n(x+z)U_k(x+u)=
\frac{1}{a}\exp\left[\frac{iy(z+u)}{2}+i(n-k)\arctan\frac{y}{z-u}\right]I_{nk}(\rho),
\label{intg} \ee where the function
$I_{nk}(\rho)=\exp(-\rho/2)L_n^{n-k}(\rho)$ is the Laguerre
function of the argument \bb \rho=\frac{y^2+(u-z)^2}{2},
\label{argg} \ee  $L_n^{n-k}(\rho)$ is the Laguerre polynomial and
in our problem $n=k=0$.

Integration over $p_3$ \bb I=\int\limits_{-\infty}^{\infty}dp_3
\frac{e^{-ip_3z/\hbar}}{p_3^2+2m|E_0|} \label{intz} \ee can be
carried out in the complex plane, closing the integration contour
into the lower half-plane for $z>0$ and in the upper half-plane
for $z<0$. The case $\lambda\ll a$ is of physical interest. For
this case, one obtains \bb I= \frac{e^{-\sqrt{2m|E_0|}|z|/\hbar}}
{\sqrt{2m|E_0|)}}. \label{intz1} \ee  Then, simple calculations
leads to the normalized electron wave function in the form (see,
also Refs. \cite{chib,TeBag,pg1}) \bb \Psi_0({\bf
r})\sim\frac{\hbar}{\sqrt{2\pi l}a}
\exp\left(-\frac{x^2+y^2-2ixy}{4a^2}\right)
\exp\left(-\frac{|z|}{l}\right),\quad
l=\frac{\hbar}{\sqrt{2m|E_0|}}. \label{ground} \ee

It is of great interest that there is the probability current even
when the electron is in the bound state in a superposition of a
constant uniform magnetic field and a single attractive
$\lambda\delta({\bf r})$ potential. The probability current
density is  \bb J_x=-\frac{\hbar }{2\pi l
ma^4}y\exp\left(-\frac{x^2+y^2}{2a^2}-\frac{2\sqrt{2}|z|}{l}\right)\equiv
-J_0y\exp\left(-\frac{x^2+y^2}{2a^2}-\frac{2\sqrt{2}|z|}{l}\right),\nonumber\\
J_y=J_0x\exp\left(-\frac{x^2+y^2}{2a^2}-\frac{2\sqrt{2}|z|}{l}\right),
\quad J_z=0.\phantom{mmmmmmm} \label{curr}\ee

We see from Eq. (\ref{curr}) that  the  divergence of the
probability current density is equal to zero everywhere.
Therefore, the probability density $|\Psi_0({\bf r})|^2$ and the
current density (\ref{curr}) satisfy  the continuity equation \bb
\frac{\partial|\Psi_0({\bf r})|^2}{\partial t}+ \frac{\partial
J_x}{\partial x}+\frac{\partial J_y}{\partial y}+\frac{\partial
J_z}{\partial z} \equiv \frac{\partial|\Psi_0({\bf
r})|^2}{\partial t}+ (\mathbf{\nabla\times J})=0 \label{cont}\ee
everywhere and  the function $|\Psi_0({\bf r})|^2$ is conserved in
time.

The electron wave functions in an external electromagnetic field
are known to have the ambiguity, which is related to the ambiguity
of the 4-potential ($\Phi, {\bf A}$) of electromagnetic field.
The latter is determined just with the exactness up to the gauge
transformation \bb \Phi\to \Phi -\frac{1}{c}\frac{\partial
f}{\partial t}, \quad {\bf A}\to {\bf A}+\mathbf{\nabla A},
\label{gauvp}\ee where $f$ is an arbitrary function of coordinates.
The Schr\"odinger equation does not change provided
the replacement of the vector potential in the Hamiltonian is
carried out simultaneously with the replacement of wave function
accordingly \bb \Psi\to \Psi\exp\left(\frac{ie}{\hbar c}f\right)
\label{gauwf}\ee.

In the considered case, we can cancel the phase factor in wave
function (\ref{ground}) by means of the gauge transformation of
vector potential with function $f=Bxy/2$. Under such a
transformation  the vector potential transforms from Eq.
(\ref{e1}) to \bb {\bf A}=\frac{B}{2}(-y, x, 0),\label{gtvp}\ee
the wave function (\ref{ground}) becomes real but the (electric)
current density \bb {\bf
J}=\frac{ie\hbar}{2m}[(\mathbf{\nabla}\Psi^*)\Psi-\Psi^*\mathbf{\nabla}\Psi]-
\frac{e^2}{mc}{\bf A}\Psi^*\Psi, \label{curmf}\ee being by the
gauge invariant quantity, does not change. Note that \bb {\bf J}=-
\frac{e^2}{mc}{\bf A}\Psi^*\Psi \label{curmf1}\ee at the gauge
(\ref{gtvp}).

The probability current density is shown in Fig. 3 in which the
coordinates $x, y, z$ are measured in units of $a$.

\begin{figure}[h]
\vspace{-0.2cm} \centering
\includegraphics[scale=1]{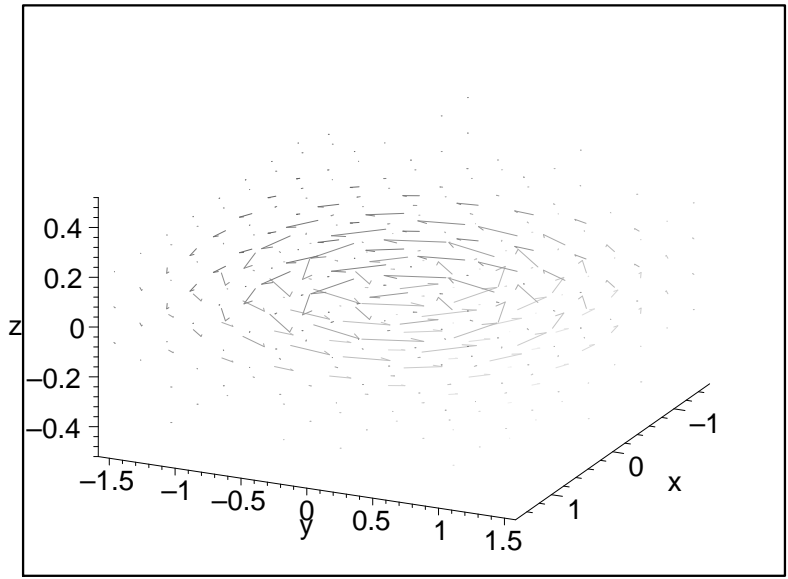}
\caption{Probability current density in space} \vspace{-0.1cm}
\end{figure}

The vector field ${\bf J}$ can be treated semiclassically. Let the
two-dimensional vector field ${\bf j}$ be a complex quantity ${\bf
j}=j_x+ij_y$ in any plane $z=constant$ whose the components $j_x$
and $j_y$ are functions of the complex variable $v=x+iy$ and
depend on $z$. The vector field ${\bf j}(x, y)$ is a vortex field.
Computing ${\bf C}=[\mathbf{\nabla\times j}]$, we obtain \bb
[\mathbf{\nabla\times j}]=\frac{\partial j_y}{\partial
x}-\frac{\partial j_x}{\partial
y}=\frac{A}{\pi}\left(1-\frac{vv^*}{2a^2}\right)\exp\left(-\frac{vv^*}{2a^2}\right),
\label{rot} \ee where $A=e\hbar F(z)/lma^4$, the function
$$
F(z)= \exp\left(-\frac{2|z|}{l}\right)
$$
should be calculated  on the plane $z=constant$  and $v^*$ is the
complex conjugate of $v$. In each plane $z=constant$ the vector
${\bf j}(x, y)$ determines the vector field of a point-like vortex
located at the point $x, y=0$.

One can write ${\bf j}$ in the form \bb {\bf j}=\frac{i I(v)}{2\pi
v^*},\label{inten} \ee where \bb
I(v)=Avv^*\exp\left(-\frac{vv^*}{2a^2}\right)\label{int1} \ee is
the vortex intensity that is the vector field circulation  ${\bf
j(x, y)}$ on any closed contour encircling the vortex at the point
$x, y=0$.

The normalized two-dimensional vector field ${\bf j}(x, y)$ is
shown in the plane $z=0$ in Fig. 4 in which the coordinates $x, y$
are measured in units of $a$.

\begin{figure}[h]
\vspace{-0.2cm} \centering
\includegraphics[angle=270, scale=0.5]{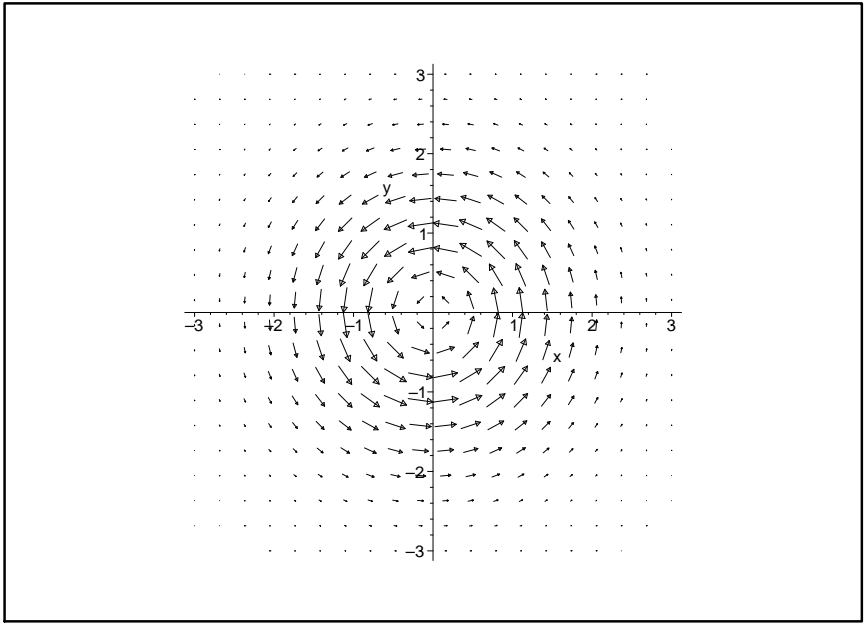}
\caption{Probability current density in the $z=0$ plane}
\vspace{-0.1cm}
\end{figure}

One can see from Eq. (\ref{curr}) that the probability current
density resembles a stack of ``pancake" vortices'' whose
circulating ``currents'' around the magnetic field direction
($z$-axes) are mostly concentrated within the plane $z=0$.

\section{Effect of Homogeneous Electric Field on the Bound Electron}

Now we consider the effect of a weak constant homogeneous electric
field on an electron bound by an attractive singular potential in
the presence of a constant uniform magnetic field. Related
problems are the ionization of negative charged ions in
accelerators as well as the removal of electrons from trapping
potentials by a constant electric field. It is worthwhile to note
that the problem concerning the removal of a charged
nonrelativistic particle from a spherically symmetric potential
well by a constant homogeneous electric field was first partly
solved in \cite{kpp} and the more rigorous formulas was given in
\cite{dd} for the total and in \cite{nr} for the differential
probability. Here we consider this problem for the model studied
with the inclusion of a weak constant homogeneous electric field
specified by the potential \bb U(z)=-|e|\epsilon z, \label{potel1}
\ee where $\epsilon>0$ is the electric field strength.

The inclusion of the electric field in the model under discussion
leads to the quasistationary of the bound electron state and the
appearance of the nonzero particle  flux at large distances from
the potential well. It is apparent that for considered
electromagnetic field combination we need calculate only the
particle flux in the $z$ direction far from the potential well.

Let us find the decay probability of the bound state per unit
time, which is equal to the electron flux across the plane
perpendicular to the $z$ axis. To find this flux we need the
Schr\"odinger equation solutions for an electron far from the
potential well. So far as the electron flux is of interest in the
$z$ direction the magnetic field cannot be taken into account far
from the well. Solutions of the Schr\"odinger equation for an
electron in field (\ref{potel1}) are expressed via the Airy
functions (see, for example, \cite{nr}).

So far as the particle flux has to be nonzero in the $z$ direction we must
choose the solution with the complex Airy function for the real $z$:
\bb
 \psi_{p_1,p_2,E}= e^{-iEt/\hbar}e^{i(p_1x+p_2y)/\hbar}V(Z),
\label{solel} \ee where \bb
Z=\left(\frac{2m|e|\epsilon}{\hbar^2}\right)^{1/3}\left(
\frac{p_1^2+p_2^2}{2m|e|\epsilon} -\frac{E}{|e|\epsilon}-z\right),
\label{argZ} \ee  $p_1, p_2$ are the eigenvalues of the operator of
generalized momentum and $V(Z)$ is  the complex Airy function
for the real $Z$. For $Z\ll 1$ \bb
 V(z) = \frac{\sqrt{\pi}}{z^{1/4}}\left[\exp\left(\frac{2z^3/2}{3}\right)
+\frac{i}{2}\exp\left(-\frac{2z^3/2}{3}\right)\right].
\label{asympel} \ee

Note that only the keeping of exponentially small term in solution
(\ref{asympel}) will give rise to the nonzero flux.

Outside the well the electron wave function  for large $z$ can be
written as the superposition of solutions in electric field with
$E=-E_0$ \bb
 \psi({\bf r})= e^{-iE_0t/\hbar}\int dp_1dp_2 e^{i(p_1x+p_2y)/\hbar} V(Z)F(p_1 p_2)
\label{wellel} \ee The appearance probability of electron far from
the well in unit time  $w$ is equal to the flux across the plane
perpendicular to the $z$ axes \bb w=-\frac{i\hbar}{2m}\int dxdy
\left(\psi^*\frac{\partial\psi}{\partial z} -
\psi\frac{\partial\psi^*}{\partial z}\right). \label{flux} \ee The
function $F(p_1 p_2)$ is found for large $z$ from Eq.
(\ref{wellel}) by the Fourier transform   \bb F(p_1
p_2)=\frac{1}{4\pi^2V(Z)}\int dxdy \psi({\bf r})
e^{i(E_0t-p_1x-p_2y)/\hbar}. \label{furie} \ee In weak electric
field  $\epsilon\ll m^{1/2}|E_0|^{3/2}/\hbar|e|\equiv \epsilon_0$
the electron wave function (\ref{wellel}) differs insignificantly
from the wave function of bound state (\ref{ground}) (we  just
consider this case) for $|z|\ll
\sqrt{\hbar^2\epsilon_0/2m|E_0|\epsilon}$, so for such $z$ we can
substitute the function (\ref{ground}) instead of $\psi({\bf r})$
in Eq. (\ref{furie}).

Carrying out simple calculations, we obtain  \bb F(p_1
p_2)=\frac{C\exp(-i-|z|/l)}{4\pi^2V(Z)}\int dxdy
e^{-\frac{x^2+y^2-2ixy}{4a^2}+i\frac{p_1x+p_2y}{\hbar}}\approx
\frac{Ca^2\exp(-|z|/l)}{\pi
V(Z)}e^{-\frac{(p_1^2+p_2^2)a^2}{\hbar^2}}, \label{furie1} \ee
where  $C=1/(a\sqrt{2\pi l})$ is the normalization constant of the
wave function of bound state. Finally, for the total probability
one obtains \bb w=-\frac{2\sqrt{\pi}\hbar
a^2|e|\epsilon}{l^2a^2m|e|\epsilon+l\hbar^2}\left(1+
\frac{ml^3|e|\epsilon}{2(a^2ml|e|\epsilon+h^2)}\right)e^{-\frac{2\hbar^2}
{3ml^3|e|\epsilon}}. \label{flux1} \ee

\section{Resume}

It is of interest to compare  wave function (\ref{ground}) with
the electron wave function of bound state in the only singular
attractive $\hbar^2\lambda\delta({\bf r})/2m$ potential. The
latter can be easily obtained in the form \bb\Psi({\bf
r})=\sqrt{\frac{1}{2\pi l_0}}\frac{e^{-r/l_0}}{r} \label{ground1}
\ee where $l_0=\sqrt{\hbar^2/2m|E_0|}$ and $E_0<0$ is the electron
energy and the wave function (\ref{ground1}) is normalized as
follows \bb\int|\Psi({\bf r})|^2dV=1. \label{ground2} \ee

The distribution of probabilities of different coordinates of the
electron
$$
|\Psi({\bf r})|4\pi r^2dr=2\exp\left(-\frac{2r}{l_0}\right)
\frac{dr}{l_0}
$$
is spherically symmetrical. Here $r=\sqrt{x^2+y^2+z^2}$. One sees
that without magnetic field the electron is localized in the
region $\sim l_0$ and the probability current density for the
state (\ref{ground1}) is equal to zero everywhere.

The distribution of probabilities of different coordinates of the
electron in the state (\ref{ground}
$$
|\Psi_0({\bf r})|2\pi \rho d\rho dz=\frac{\rho}{a}
\exp\left(-\frac{\rho^2}{2a^2}-\frac{2|z|}{l}\right)
\frac{d\rho}{a}\frac{dz}{l}
$$
has the cylindrical symmetry. Here $\rho=\sqrt{x^2+y^2},\quad z$
are the cylindrical coordinates. The electron is located in the
region $x, y\sim \sqrt{2}a$ in the $x,y$ plane perpendicular to
the magnetic field and $z\sim 2l$ along the magnetic field
direction.

It is of importance that the probability current arises even when
the electron is in the bound state a superposition of a constant
uniform magnetic field and a single attractive $\lambda\delta({\bf
r})$ potential. This (electric) current is the gauge invariant
(physically observable) quantity and, therefore, it must involve
(and exercise influence on) the electron scattering.

\vspace{0.5cm} \noindent{\bf Acknowledgments}

This paper was supported  by a Joint Research Project  of the
Taiwan National Science Council (NSC-RFBR No. 95WFD0400022,
Republic of China) and the Russian Foundation for Basic Research
(No. NSC-a-89500.2006.2) under Contract No. RP06N04-1, by  the
U.S. Department of Energy's Initiative for Proliferation
Prevention (IPP) Program through Contract No. 94138 with the
Brookhaven National Laboratory, and, in part, by the Program for
Leading Russian Scientific Schools (Grant No.
NSh-5332.2006.2)(V.R. K.).

\end{document}